\documentclass[aps,preprint,tightenlines,superscriptaddress]{revtex4}
\usepackage{epsfig}
\usepackage{graphicx}
\usepackage{psfrag}
\usepackage{amsmath,amssymb}
\usepackage{colordvi}
\usepackage{amsfonts}
\usepackage{enumerate}
\usepackage{slashed}
\usepackage{multirow}
\usepackage{color}

\graphicspath{{ChPT_Graphs/}}

\begin{document}
\title{Direct CP Violation in K-decay \\ and Minimal Left-Right Symmetry Scale}
\author{Panying Chen}
\affiliation{Department of Physics, University of Maryland, College
Park, Maryland 20742, USA}
\author{Hongwei Ke}
\affiliation{Department of Physics, University of Maryland, College
Park, Maryland 20742, USA}
\affiliation{Department of Physics, Nankai University, Tianjin, 300071, P. R. China}
\author{Xiangdong Ji}
\affiliation{Department of Physics, University of Maryland, College
Park, Maryland 20742, USA} \affiliation{Center for High-Energy
Physics and Institute of Theoretical Physics, \\ Peking University
Beijing, 100080, P. R. China}

\date{\today}
\vspace{0.5in}
\begin{abstract}
We calculate the new contribution to the direct CP-violation parameter
$\epsilon'$ in $K\rightarrow \pi\pi$ decay in the minimal left-right symmetric model
with the recently-obtained right-handed quark Cabibbo-Kobayashi-Maskawa mixing .
We pay particular attention to the uncertainty in the hadronic matrix
element of a leading four-quark operator $O^{LR}_-$. We find that it can be related
to the standard model electromagnetic penguin operator $O_8$ through
$SU(3)_L \times SU(3)_R$ chiral symmetry. Using the lattice and large $N_c$
calculations, we obtain a robust constraint on the minimal
left-right symmetric scale $M_{W_R}>5$ TeV from the experimental data on $\epsilon'$.
\end{abstract}
\maketitle

One of the much studied themes for particle physics beyond the standard model (SM)
is left-right symmetry at high-energy, introduced many years ago by Mohapatra and Pati ~\cite{lrmodel}. In a recent work, it has been shown that supersymmetric left-right
theory arises naturally from duality cascade of a quiver in the context of intersecting D-branes~\cite{Heckman:2007zp}. The twin-Higgs model, introduced to explain the
disparity between the new physics scale and the electroweak scale \cite{twinhiggs},
also utilizes the idea of left-right symmetry. However,
the direct collider search for the signatory right-handed $W$ gauge boson shows
that it is at least 10 times heavier than its left-handed counterpart~\cite{PDG}.
The most stringent limit on the right-handed scale
has been obtained from low-energy data, with the most well-known example being the neutral kaon mass
difference~\cite{soni}, which gives a lower bound of at least $2.0-2.5$ TeV.

More recently, a general solution for the right-handed Cabibbo-Kobayashi-Maskawa (CKM)
quark mixing in the minimal left-right symmetric model (LRSM) has been found~\cite{Zhang:2007fn}. Particularly interesting is the
CP(charge-conjugation-parity)-violating mechanisms in the model: Apart from the usual
Dirac CP phase appearing in the left-handed CKM mixing, there is also a spontaneous symmetry-breaking phase $\alpha$ that contributes to CP-violating observables.
Using the neutral kaon mixing parameter $\epsilon$, $\alpha$ can be constrained
accurately. Therefore, one can make predictions on other
CP-violating observables including the neutron electrical dipole moment (EDM)
and direct CP-violating parameter $\epsilon'$ in kaon decay; the experimental
data can then provide new constraints on the left-right symmetric scale~\cite{Zhang:2007da}.
Unfortunately, the intermediate steps involve unknown hadronic matrix elements, and the simple factorization or large $N_c$ (number of quark colors) assumption is usually
adopted to make estimations in previous studies~\cite{Zhang:2007da,Ecker:1985vv}. As a consequence,
the bounds suffer from unknown hadronic physics uncertainties,
as exemplified in reproducing the $\Delta I=1/2$ rule for the $K$ to $\pi\pi$ decay.

In this paper, we focus on a better estimation of the uncertainty associated
with the leading hadronic matrix element, and hence a more accurate bound on the
minimal left-right symmetry scale. In particular, we have found a relation between
the dominating four-quark operator $O^{LR}_-$ in the new contribution and the SM electromagnetic
penguin operator $O_8$ through $SU(3)_L\times SU(3)_R$ chiral symmetry. We use the
existing knowledge on the matrix element of the latter to get information on
the former~\cite{Buras:2003zz}. With a reasonable estimate of the $O^{LR}_-$
matrix element, we find the lower bound for the right-handed scale in the
range of 5-9 TeV, consistent with that from the neutron EDM data~\cite{Zhang:2007da}.

The direct CP-violation parameter in the neutral kaon to $\pi\pi$ decay is calculated via
\begin{equation}
\epsilon'=\frac{i}{\sqrt{2}} \omega \left(\frac{q}{p}\right) \left(\frac{{~\rm{Im}}A_2}
{{~\rm{Re}}A_2}-\frac{{~\rm{Im}}A_0}{{~\rm{Re}}A_0}\right) e^{i(\delta_2-\delta_0)} \ ,
\end{equation}
where the decay amplitudes $A_0$ and $A_2$  are defined as the
matrix elements of the $\Delta S = 1$ effective Hamiltonian between
the neutral-K meson and the isospin $I= 0$ and $2$ $\pi\pi$ states,
\begin{equation}
\langle(2\pi)_I|(-i){\cal H}_{\Delta S=1}|K^0\rangle=A_Ie^{i\delta_I}  \ .
\end{equation}
$\delta_I$ is the strong phase for $\pi\pi$ scattering at the kaon mass,
$\omega \equiv A_2 / A_0$, and
$p$, $q$ are the mixing parameters for $K^0-\overline{K}^0$. To an excellent
approximation, $\omega$ can be taken as real and $q/p = 1$. We use the
experimental value for the real parts of $A_0$ and $A_2$: Re$ A_0 \simeq 3.33\times
10^{-7}$~GeV and $\omega \simeq 1/22$. We focus on calculating the imaginary
part of the decay amplitudes.

In the SM, the contributions to $\epsilon'$ come from both QCD and
electromagnetic penguin diagrams~\cite{Shifman:1976ge}. The QCD
penguin contributes exclusively to the imaginary part of $\Delta I =1/2$ decay, whereas
the electromagnetic penguin is mainly responsible for the imaginary part of $\Delta I
=3/2$ decay. Both contributions are important but have opposite
signs. Therefore, the final result depends on delicate cancelations
of hadronic matrix elements. The state-of-art chiral perturbation
theory~\cite{Buras1,Buchalla:1989we,prime,Bosch:1999wr} and lattice
QCD calculations~\cite{Blum:2001xb,Pekurovsky:1998jd} have not yet
been sufficiently accurate to reproduce the experimental result~\cite{expprime}.
On the other hand, a large-$N_c$ approach with final-state rescattering
effect taken into account seems to be able to reproduce
the experimental result~\cite{pich}. A nice review of the SM calculation can be
found in Ref.~\cite{Bertolini:2000dy,Buras:2003zz}.

\begin{figure}[hbt]
\begin{center}
\includegraphics[width=10cm]{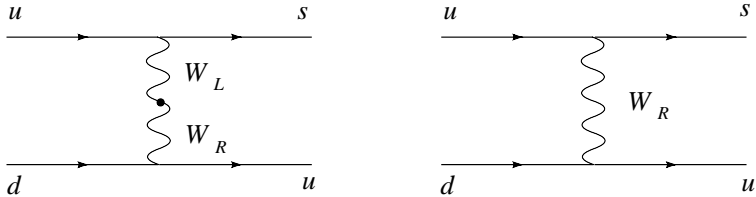}
\end{center}
\caption{New tree-level contributions to the $\Delta S = 1$ interaction from LRSM.}
\label{e'}
\end{figure}

In LRSM, every element in the right-handed CKM matrix has a substantial CP phase. As a
consequence, there are tree-level contributions to the phases of $A_2$ and $A_0$.
Following closely the work by Ecker and Grimus \cite{Ecker:1985vv}, the tree-level Feynman diagrams in Fig. \ref{e'} generate
\begin{widetext}
\begin{eqnarray}\label{H}
{\cal H}^{\rm tree}_{\Delta S =1}
&=&\frac{G_F}{2\sqrt{2}}\lambda^{LL}_u\left[\left(\frac{\alpha_S(\mu^2)}
{\alpha_S(M^2_L)}\right)^{-\frac{2}{b}}O^{LL}_+(\mu) +\left(\frac{\alpha_S(\mu^2)}
{\alpha_S(M^2_L)}\right)^{\frac{4}{b}}O^{LL}_-(\mu)\right]\nonumber\\
&+& \frac{G_F}{2\sqrt{2}}\frac{M^2_L}{M^2_R}\lambda^{RR}_u\left[\left(\frac{\alpha_S(\mu^2)}
{\alpha_S(M^2_R)}\right)^{-\frac{2}{b}}O^{RR}_+(\mu) +\left(\frac{\alpha_S(\mu^2)}
{\alpha_S(M^2_R)}\right)^{\frac{4}{b}}O^{RR}_-(\mu)\right]\nonumber\\
&+&\frac{G_F}{\sqrt{2}}\sin\zeta\lambda^{LR}_ue^{i\alpha}\left[\left(\frac{\alpha_S(\mu^2)}
{\alpha_S(M^2_L)}\right)^{-\frac{1}{b}}O^{LR}_+(\mu) -\left(\frac{\alpha_S(\mu^2)}
{\alpha_S(M^2_L)}\right)^{\frac{8}{b}}O^{LR}_-(\mu)\right]\nonumber\\
&+&\frac{G_F}{\sqrt{2}}\sin\zeta\lambda^{RL}_ue^{-i\alpha}\left[\left(\frac{\alpha_S(\mu^2)}
{\alpha_S(M^2_L)}\right)^{-\frac{1}{b}}O^{RL}_+(\mu) -\left(\frac{\alpha_S(\mu^2)}
{\alpha_S(M^2_L)}\right)^{\frac{8}{b}}O^{RL}_-(\mu)\right] \ ,
\end{eqnarray}
\end{widetext}
where we have taken into account the leading-logarithm QCD corrections with
renormalization scale $\mu$ taken to be around the charm quark mass $m_c \sim$ 1.3 GeV,
and $b = 11-2N_f/3$ with $N_f$ the
number of active fermion flavors. The left-right mixing parameter is
\begin{equation}
    \tan \zeta = 2r \frac{m_b}{m_t} \left( \frac{M_{W_L}}{M_{W_R}}\right)^2 \ ,
\end{equation}
where $r$ is a parameter less than 1.
The mixing coupling $\lambda^{AB}_u=V^{\rm
CKM*}_{Aus}V^{\rm CKM}_{Bud}$, $A$, $B$ are $L$, $R$. The right-handed CKM matrix has a form,
\begin{eqnarray}\label{vr}
V_R = P_U \widetilde V_L P_D \ ,
\end{eqnarray}
in which $P_U={\rm diag}(s_u,s_ce^{2i\theta_2},s_te^{2i\theta_3})$, $P_D={\rm
diag}(s_de^{i\theta_1},s_se^{-i\theta_2},s_be^{-i\theta_3})$, and
\begin{equation}\label{vr1}
\widetilde V_L=\left(\begin{array}{ccc} 1-\lambda^2/2 & \lambda &A\lambda^3(\rho-i\eta)\\
-\lambda & 1-\lambda^2/2 &
A\lambda^2e^{-2i\theta_2}\\A\lambda^3(1-\rho-i\eta)&-A\lambda^2e^{2i\theta_2}&1\\
\end{array}\right) \ ,
\end{equation}
where $\lambda$, $A$, $\rho$ and $\eta$ are Wolfenstein parameters and the new
phases $\theta_i$ are all related to spontaneous CP phase $\alpha$,
\begin{eqnarray}\label{theta}
\theta_1&=&-\sin^{-1}[0.31(s_ds_c+0.18s_ds_t)r\sin\alpha]\ , \nonumber\\
\theta_2&=&-\sin^{-1}[0.32(s_ss_c+0.25s_ss_t)r\sin\alpha]\ , \nonumber\\
\theta_3&=&-\sin^{-1}[s_bs_tr\sin\alpha] \ ,
\end{eqnarray}
where experimental quark masses have been used with possible $s_i=\pm 1$ signs.
The four-quark operators are
\begin{eqnarray}\label{O} O^{LL,RR}_{\pm}&=&(\overline{s}_i u_i)_{V\mp A}
(\overline{u}_j d_j)_{V\mp A} \pm (\overline{s}_i  d_j)_{V\mp A} (\overline{u}_j  u_j )_{V\mp A}\ , \nonumber\\
O^{LR,RL}_+&=& (\overline{s}_i u_i)_{V\mp A} (\overline{u}_j  d_j)_{V\pm A}-\frac{1}{3}
(\overline{s}_i u_j)_{V\mp A}(\overline{u}_j d_i)_{V\pm A} \ , \nonumber\\
O^{LR,RL}_-&=&-\frac{1}{3}(\overline{s}_i  u_j)_{V\mp A} (\overline{u}_j d_i)_{V\pm A}
\ ,
\end{eqnarray}
where $i$ and $j$ are color indices and the subscript $V\pm A$ refers to a quark bilinear
of the form $\bar q\gamma_\mu (1\pm \gamma_5) q$.

As mentioned above, one has to include the penguin contributions in the SM calculation because
the CKM matrix elements have non-zero CP phases only when the third family is introduced.
The only detail we would like to point out about the SM contribution is that
the electromagnetic penguin involves predominantly the following operator
\begin{equation}
  O_8 = \frac{1}{2} (\bar s_i d_{j})_{A-V} \left[2 (\bar u_{j}u_{i})_{V+A} -  (\bar d_{j} d_{i})_{V+A}
        - (\bar s_{j} s_{i})_{V+A}\right] \ ,
\end{equation}
which is an $(8,8)$ representation of the chiral $SU(3)_L\times SU(3)_R$ group.
In principle, there are also new QCD penguin diagrams
involving the right-handed gauge boson, particularly with left-right
gauge boson mixing. However, these contributions are suppressed by a loop factor
relative to the tree contributions as well as the $\Delta I=1/2$ rule, and
hence are neglected \cite{Ecker:1985vv}.

Now we come to estimate the new contributions to the direct CP-violation parameter $\epsilon'$.
There are two types of tree contributions: the right-handed current alone and left-right
interference. Both are nominally the same size, and are suppressed by
$1/M_{W_R}^2$ relative to the SM contribution. In practice, however, the interference
contribution dominates numerically. Let us consider the right-handed current
contribution first. The relevant hadronic matrix elements can be obtained from
the SM ones through parity transformation,
\begin{equation}
     \langle \pi\pi | O^{RR}_\pm |K_0 \rangle = - \langle \pi\pi | O^{LL}_\pm |K_0 \rangle \ .
\end{equation}
We use the matrix elements from a domain-wall lattice QCD calculation~\cite{Blum:2001xb},
which are consistent with the $\Delta I=1/2$ rule,
\begin{eqnarray}
  && \langle (\pi\pi)_{I=0} | O^{LL}_- |K_0 \rangle = 0.192i~ {\rm GeV}^3 \nonumber \\
  && \langle (\pi\pi)_{I=0} | O^{LL}_+ |K_0 \rangle = 0.064i~ {\rm GeV}^3 \nonumber \\
  && \langle (\pi\pi)_{I=2} | O^{LL}_+ |K_0 \rangle = 0.025i~ {\rm GeV}^3
\end{eqnarray}
The matrix element of $O^{LL}_+$ in $I=0$ state is less important and can largely be
ignored.

The dominating new contribution is from the left-right W-boson interference.
Due to the QCD running effect and chiral suppression, $O_+^{LR}$ operator is less important
relative to $O_-^{LR}$ and hence will be ignored. Therefore, we need to consider only
the matrix element of $O_-^{LR}$ operator in the $I=2$ state. Introduce the
following $(8,8)$ operators,
\begin{eqnarray}
O_{3/2}^{(8,8)} &=& (\bar s_i d_j)_{V-A} (\bar u_ju_i)_{V+A} + (\bar s_i u_j)_{V-A} (\bar u_jd_i)_{V+A}
  \nonumber \\ &-& (\bar s_i d_j)_{V-A} (\bar d_jd_i)_{V+A} \ ,  \nonumber \\
O_{1/2A}^{(8,8)} &=& (\bar s_i d_j)_{V-A} (\bar u_ju_i)_{V+A} - (\bar s_i u_j)_{V-A} (\bar u_jd_i)_{V+A}
  \nonumber \\  &-& (\bar s_i d_j)_{V-A} (\bar s_j s_i)_{V+A} \ , \\
O_{1/2S}^{(8,8)} &=& (\bar s_i d_j)_{V-A} (\bar u_ju_i)_{V+A} + (\bar s_i u_j)_{V-A} (\bar u_jd_i)_{V+A} \nonumber \\
  &+& 2(\bar s_i d_j)_{V-A} (\bar d_j d_i)_{V+A}- 3(\bar s_i d_j)_{V-A} (\bar s_js_i)_{V+A}
  \nonumber \ ,
\end{eqnarray}
where subscripts 3/2 and 1/2 indicate isospin. Using the above, one can express $O_-^{LR}$ as follows
\begin{equation}
O_-^{LR} = - \frac{1}{9} O^{(8,8)}_{3/2} - \frac{1}{18} O^{(8,8)}_{1/2S} + \frac{1}{6}O^{(8,8)}_{1/2A} \ .
\end{equation}
On the other hand, the electromagnetic penguin operator $O_8$ can be expressed as
\begin{equation}
  O_8 = \frac{1}{2} \left(O^{(8,8)}_{3/2} + O^{(8,8)}_{1/2A}\right) \ .
\end{equation}
Therefore, we find the model-independent relation,
\begin{equation}
  \langle (\pi\pi)_{I=2}| O_-^{LR} |K_0\rangle
   = - \frac{2}{9}    \langle (\pi\pi)_{I=2}| O_8|K_0\rangle \ .
\end{equation}

\begin{figure}[hbt]
\begin{center}
\includegraphics[width=8cm]{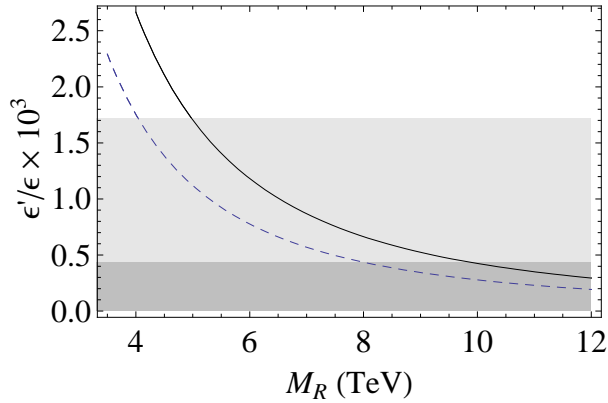}
\end{center}
\caption{The new contribution in LRSM to $\epsilon'$ as a function of $M_{W_R}$ for $\sin \alpha =
0.1$, $r=0.5$ with $s_d=s_s=-1$ and all other $s_q =1$.
The light shaded part is allowed by the experimental data, and the heavy-shaded area is
1/4 of the experimental data.}\label{e'2} \end{figure}

In the vacuum insertion approximation, one finds
\begin{equation}
  \langle (\pi\pi)_{I=2}|O_8||K_0\rangle = \sqrt{6} f_\pi \left(\frac{m_K^2}{m_s(\mu)+m_d(\mu)}\right)^2 i \ ,
\end{equation}
which is about $0.95i$ GeV$^3$ ($f_\pi = 93$ MeV) if the strange quark mass is taken to be 120 MeV at the scale of $m_c$. On the other hand, the lattice QCD calculation in
Ref.~\cite{Blum:2001xb} gives $1.4i$ GeV$^3$ at the scale of
1.9 GeV. This lattice calculation, however, does not reproduce the experimental data
on $\epsilon'$. In Ref. \cite{Buras:2003zz}, an extensive discussion has been made
about the size of this matrix element. It is expected that the variation of the matrix
element is between 1 to 2 of the factorization result.

Because the phase $\alpha$ in the factor $e^{i\alpha}$ is dominating,
$\epsilon'$ is approximately a function of $r\sin \alpha$, rather than
$r$ and $\sin \alpha$ independently. Since $r \sin \alpha$ has
been fixed by $\epsilon$ and neutron EDM $d_n^e$ \cite{Zhang:2007da}, $\epsilon'$ is
approximately a function of $M_{W_R}$ only. In Fig. \ref{e'2}, we plot $\epsilon'$ as a
function of $M_{W_R}$ for $\sin \alpha = 0.1$,
$r=0.5$ and $s_ds_s =1$ which is required by the neutron EDM calculation.
[All other $s_i = 1$.] We choose the renormalization scale at the charm quark mass
and $\Lambda_{\rm QCD}=340$ MeV. The dashed curve shows the result with the large-$N_c$
matrix element, whereas the solid curve shows that from the lattice QCD~\cite{Blum:2001xb}.

If one uses that factorized matrix element and following the Refs.
\cite{pich,Buras:2003zz} for other hadronic matrix elements,
the experimental data is roughly reproduced by the SM calculation.
Requiring the new contribution is less than 1/4 of the
experimental data, we get a large lower bound of 8 TeV on the right-handed
scale. On the other hand, if one takes the calculation in Ref. \cite{Blum:2001xb}
seriously, the lattice QCD generates a small and negative contribution
to $\epsilon'$. If then requiring that the experimental number is entirely
reproduced by the new contribution, we find a limit on $M_{W_R}$ about 5 TeV.
In any case, $\epsilon'$ gives a tighter lower bound on $M_{W_R}$ than the well-known
neutral kaon mass difference. If on the other hand, we take $r\sin\alpha=0.15$,
as required by low $M_H$, the bound changes to 8.5 TeV. Therefore, we take the
range 5-8 TeV as our final estimate.

Finally, we have also calculated the tree-level flavor-changing neutral Higgs contributions
to ${\cal H}_{\Delta S=1}$. Since the relevant coupling is suppressed by either
the Cabibbo angle or the quark masses, their contribution is negligible.

To conclude, we have found that a robust bound on the mass of the right-handed $W$-boson
based on a relatively well-known estimate on the strong interaction matrix element of
$O_-^{LR}$, which is known to within a factor of 2.
The result is on the order of 5-8 TeV, which is just on the border
for the Large Hadron Collider detection. This situation turns out to be better than
the similar calculation in SM.

We thank R. Mohapatra and Y. Zhang for numerous discussions related to the subject of this paper.
This work was partially supported by the U. S. Department of Energy via grant
DE-FG02-93ER-40762. H. W. Ke acknowledges a scholarship support from China's Ministry of Education.

\end{document}